\documentclass[sigconf]{acmart}
\AtBeginDocument{%
  }

\setcopyright{cc}
\setcctype{by}
\copyrightyear{2026}
\acmYear{2026}
\acmDOI{10.1145/3786583.3786875}
\acmConference[ICSE-SEIP '26]{2026 IEEE/ACM 48th International Conference on Software Engineering}{April 12--18, 2026}{Rio de Janeiro, Brazil}
\acmBooktitle{2026 IEEE/ACM 48th International Conference on Software Engineering (ICSE-SEIP '26), April 12--18, 2026, Rio de Janeiro, Brazil}
\acmPrice{}
\acmISBN{979-8-4007-2426-8/2026/04}

\usepackage[ruled,vlined,linesnumbered]{algorithm2e}
\usepackage{framed}
\usepackage{booktabs}
\usepackage{tabularx}
\usepackage{tikz}
\usepackage{pifont}     
\usepackage{wasysym}    
\newcommand{\suitable}{\ding{51}}
\newcommand{\partsuitable}{\ding{109}}
\newcommand{\notsuitable}{\ding{55}}
\usepackage{listings}
\usepackage{caption}
\lstdefinestyle{sparqlcompact}{
	language=SPARQL,
	basicstyle=\ttfamily\scriptsize,
	numbers=none,
	frame=single,
	breaklines=true,
	breakatwhitespace=false,
	columns=fullflexible,
	keepspaces=true,
	showstringspaces=false,
	upquote=true,
	xleftmargin=1em,
	aboveskip=4pt,
	belowskip=4pt,
	postbreak=\mbox{\textcolor{gray}{$\hookrightarrow$}\space}
}
\usepackage{amsmath}
\usepackage{mathtools}

\newcommand{\Aex}{A_{\mathrm{exp}}}      
\newcommand{\Aexpos}{A_{\mathrm{expo}}}  
\newcommand{\Wex}{W_{\mathrm{exp}}}
\newcommand{\Wexpos}{W_{\mathrm{expo}}}
\newcommand{\Sexp}{S_{\mathrm{exp}}}
\newcommand{\Sexpos}{S_{\mathrm{expo}}}
\newcommand{\tlow}[1]{\theta_{#1,\mathrm{lo}}}
\newcommand{\thi}[1]{\theta_{#1,\mathrm{hi}}}

\begin{document}

\title{An Ontology-Based Approach to Security Risk Identification for Container Deployments in OT Contexts}

\author{Yannick Landeck}
\email{landeck@fortiss.org}
\orcid{0009-0008-0340-3602}
\author{Dian Balta}
\email{balta@fortiss.org}
\orcid{0000-0001-8311-3227}
\affiliation{%
	\institution{fortiss GmbH}
	\city{Munich}
	\country{Germany}
}

\author{Martin Wimmer}
\email{martin.r.wimmer@siemens.com}
\orcid{0009-0009-2716-8886}
\author{Christian Knierim}
\email{christian.knierim@siemens.com}
\orcid{0000-0002-5713-4654}
\affiliation{%
	\institution{Siemens AG}
	\city{Munich}
	\country{Germany}
}

\renewcommand{\shortauthors}{Landeck, et al.}

\begin{abstract}
  In operational technology (OT) contexts, containerised applications often require elevated privileges to access low-level network interfaces or perform administrative tasks such as application monitoring.
  These privileges reduce the default isolation provided by containers and introduce significant security risks.
  Security risk identification for OT container deployments is challenged by hybrid IT/OT architectures, fragmented stakeholder knowledge, and continuous system changes.
  Existing approaches lack reproducibility, interpretability across contexts, and technical integration with deployment artefacts.
  We propose a model-based approach---implemented as the \textit{Container Security Risk Ontology (CSRO)}---which integrates five key domains: adversarial behaviour, contextual assumptions, attack scenarios, risk assessment rules, and container security artefacts.
  Our evaluation of \textit{CSRO} in a case study demonstrates that the end-to-end formalisation of risk calculation---from artefact to risk level---enables automated and reproducible risk identification. 
  While \textit{CSRO} currently focuses on technical, container-level treatment measures, its modular and flexible design provides a solid foundation for extending the approach to host-level and organisational risk factors.
\end{abstract}




\maketitle

\section{Introduction}
Software containers are increasingly adopted in operational technology (OT) contexts to leverage benefits established in information technology (IT) domains---such as flexible, scalable, and efficient application management~\cite{arnoldIIoTPlatformsArchitectural2022}.
Despite these benefits, their use in OT environments introduces new security challenges, particularly due to the elevated privileges and reduced isolation often required~\cite{martinDockerEcosystemVulnerability2018}.
These privileges are typically needed to access low-level network interfaces or perform administrative tasks such as application monitoring~\cite{miellDockerPractice2019}.
Security risk assessment becomes essential to ensure safe and reliable operation of containerised applications in these critical domains~\cite{landeckAssuranceApplicationSecurity2024}.
However, a prerequisite for effective risk assessment is comprehensive security risk identification \cite{refsdalCyberRiskManagement2015}, which remains an underexplored task in OT container deployments \cite{landeckKnowledgeAugmentedSecurityRisk2025}.

Identifying security risks for container deployments in OT contexts is challenged by the complexity of hybrid IT/OT architectures, the need for context-specific knowledge exchange, and continuous system changes~\cite{landeckSoftwareManufacturingIndustry2023, wongSecurityContainersThreat2023}.
To understand how current approaches address these challenges, we reviewed the state of the art across three categories: methods for security risk identification (e.g. CORAS~\cite{lundModelDrivenRiskAnalysis2011}), security controls for containers (e.g. AppArmor policies~\cite{zhuLicSecEnhancedAppArmor2021}), and security knowledge management (e.g. cybersecurity knowledge graphs~\cite{sikosCybersecurityKnowledgeGraphs2023}).
While these approaches offer structured frameworks, policy mechanisms, and curated resources, they fail to meet the specific demands of OT container environments—particularly in terms of reproducibility, interpretability, and technical integration.
These gaps motivate our research question:

\textbf{RQ: How can we enable reproducible, interpretable, and technically integrated security risk identification for container deployments in OT contexts?}

\begin{figure*}[htbp]
	\centering
	\includegraphics[width=\textwidth]{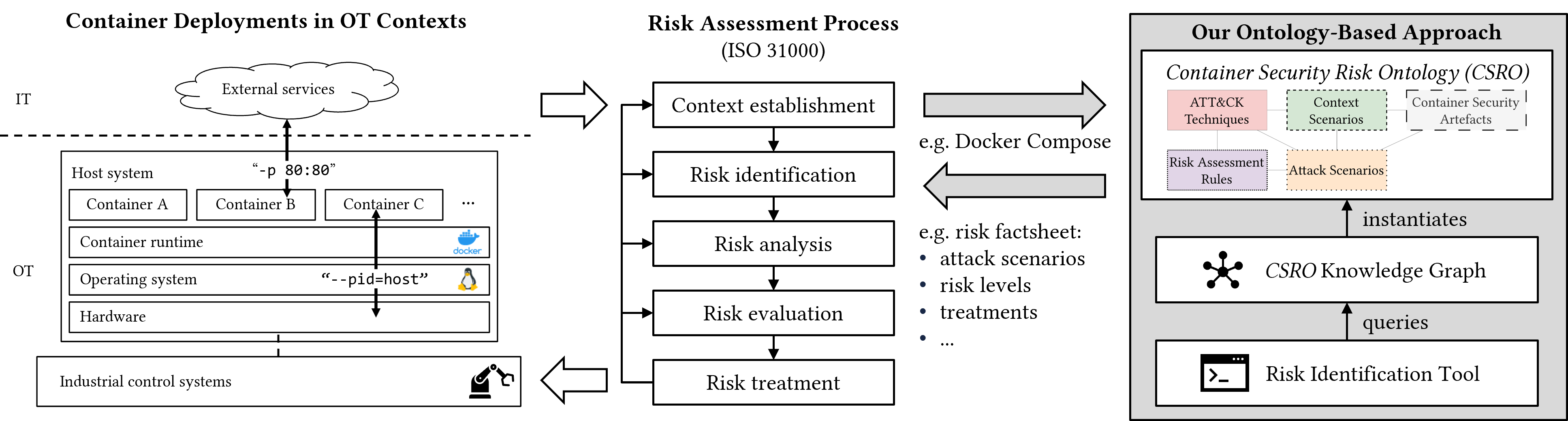}
	\caption{Overview of our ontology-based approach to complement risk assessment for container deployments in OT contexts using the \textit{Container Security Risk Ontology (CSRO)} and an integrated risk identification tool.}
	\label{fig:our_approach}
	\Description{The figure is divided into three main sections. The first section, titled "Container Deployments in OT Contexts," illustrates the interaction between IT and OT environments, showing a host system with multiple containers and external service connections. The second section, "Risk Assessment Process (ISO 31000)," outlines five steps from context establishment to risk treatment, with inputs like Docker Compose and outputs such as risk factsheets. The third section, "Our Ontology-Based Approach," presents the CSRO ontology and its integration into a knowledge graph queried by a risk identification tool.}
\end{figure*}

As a solution, we propose a model-based approach that complements risk assessment with an instantiated \textit{Container Security Risk Ontology (CSRO)} and a risk identification tool (cf. Figure~\ref{fig:our_approach}).
\textit{CSRO} formalises a semantic knowledge model that enables structured representation, automated reasoning, and integration with container security artefacts.
\textit{CSRO} comprises five key domains:
(1) Represent adversary behaviour covering Techniques from the MITRE ATT\&CK framework~\cite{stromMITREATTCKDesign2018}.
(2) Define context scenarios based on curated assumptions from container security standards (e.g. NIST Application Container Security Guide~\cite{souppayaApplicationContainerSecurity2017}).
(3) Model attack scenarios that link adversary behaviour to context scenarios.
(4) Formalise risk assessment rules to calculate the risk level for an attack scenario.
(5) Link container security artefacts such as configuration files to support automated verification and treatment derivation.

We demonstrate our approach through a case study of a large-scale industrial software platform for containerised applications in OT contexts.
We instantiate \textit{CSRO} into a knowledge graph and integrate a tool for automated security risk identification.

Our research contributes to theory by introducing an interdisciplinary, model-based approach to risk assessment that links domain-specific security knowledge with multifaceted risk representations.
We also contribute to practice by providing a structured and reproducible method for risk identification, enabling stakeholders in OT container deployments---such as container developers and system operators---to automatically derive effective treatments using \textit{CSRO}, its knowledge graph, and the integrated tool.

\section{Background and Motivation}
\subsection{Software Containers in OT Contexts}
Software containers are widely adopted for developing and deploying lightweight, flexible, and resource-efficient software applications~\cite{sultanContainerSecurityIssues2019, martinDockerEcosystemVulnerability2018, koskinenContainersSoftwareDevelopment2019}.
Compared to virtual machines, containers enable application virtualisation by sharing the operating system kernel between container processes and the host~\cite{solteszContainerbasedOperatingSystem2007, souppayaApplicationContainerSecurity2017}.
Container isolation is implemented using Linux kernel features such as namespaces, control groups (cgroups), and capabilities~\cite{wongSecurityContainersThreat2023}.

Containerised applications are increasingly deployed in operational technology (OT) contexts, e.g. on edge devices~\cite{arnoldIIoTPlatformsArchitectural2022}.
OT refers to systems used in industries such as manufacturing, transportation, or utilities, and are subject to domain-specific requirements, particularly regarding safety and security~\cite{boyesIndustrialInternetThings2018, landeckSoftwareManufacturingIndustry2023}.
Container technologies are adopted in OT to accelerate development, enhance flexibility, and enable use cases such as external production management, predictive maintenance, and anomaly detection~\cite{goldschmidtContainerbasedArchitectureFlexible2018, alcacerScanningIndustryLiterature2019, guggenbergerHowDesignIIoTPlatforms2021}.

\subsection{Importance of Security Risk Assessment} \label{sec:security_risk_assessment}
Various use cases for containers in OT contexts---such as anomaly detection---require elevated runtime privileges to transmit data, access low-level network interfaces, or perform administrative tasks~\cite{miellDockerPractice2019, bradyDockerContainerSecurity2020, arnoldIIoTPlatformsArchitectural2022}.
One approach to providing these privileges is to use deployment settings.
For instance, using Docker, the \texttt{pid=host} setting places the container in the host’s process namespace, removing isolation and granting access to the full process stack.

Deploying containers with elevated privileges introduces significant security risks in OT environments~\cite{souppayaApplicationContainerSecurity2017, martinDockerEcosystemVulnerability2018}.
Modifying the isolation between host and containers increases the likelihood and impact of threats.
Incidents like Stuxnet~\cite{chenLessonsStuxnet2011} and the Mirai botnet~\cite{nguyenAdvancedComputingApproach2022} have demonstrated the devastating consequences of security breaches in OT contexts.
Operators of OT systems therefore prioritise robust risk assessment during container deployment~\cite{landeckAssuranceApplicationSecurity2024}.
Risk assessment involves identifying potential risks within a system’s operational context, followed by their analysis, evaluation, and treatment~\cite{internationalorganizationforstandardizationISO31000Risk2018}.
Only with a clear understanding of these risks, stakeholders can implement effective countermeasures, such as access controls or tailored security policies.

\begin{framed}
	\noindent
	\textbf{Running example: shellcode injection enabled by \texttt{pid=host} and \texttt{SYS\_PTRACE} for a container.}
	
	The \texttt{pid=host} setting is often used in combination with adding the Linux capability \texttt{SYS\_PTRACE}, allowing a container to run the \texttt{ptrace(2)} syscall, e.g. to monitor other applications on the host \cite{miellDockerPractice2019, nickoloffDockerAction2019}.
	This combination poses a privilege escalation risk, as it enables an attacker to inject shellcode into a running process and escape the container \cite{ayalonContainerEscapeAll2022, hertzAbusingPrivilegedUnprivileged2016}.
	In OT contexts, such a container escape can compromise critical software components, potentially leading to severe financial losses or endangering human safety \cite{martinDockerEcosystemVulnerability2018}.
\end{framed}

\subsection{Challenges in Security Risk Identification}
A critical prerequisite for risk assessment is comprehensive security risk identification, which entails recognising threat sources, specific threats, vulnerabilities, and their associated risks~\cite{refsdalCyberRiskManagement2015}.
Identifying security risks in OT container deployments presents three technical and organisational challenges: \textit{C1--C3}.

\paragraph{C1: Complexity of Security Risk Identification}
OT container deployments introduce complexity because host system operators must assess threats from privileged applications---often developed externally or lacking transparency---and evaluate their impact in an OT context~\cite{landeckSoftwareManufacturingIndustry2024, tangeSystematicSurveyIndustrial2020}.
OT systems often prioritise availability and integrity over confidentiality, shifting the threat landscape compared to traditional IT environments~\cite{landeckSoftwareManufacturingIndustry2023, prinslooReviewIndustryManufacturing2019, boyesIndustrialInternetThings2018}.
The convergence of IT and OT introduces hybrid architectures that blur traditional security boundaries and create new dependencies and attack surfaces~\cite{landeckSoftwareManufacturingIndustry2023, dragoniMicroservicesYesterdayToday2017}.

\paragraph{C2: Need for Context-Specific Knowledge Exchange}
Security risk identification relies on stakeholders---such as developers, operators, and security experts---interpreting shared artefacts like Docker Compose files to assess risks~\cite{boliciStigmergicCoordinationFLOSS2016, landeckAssuranceApplicationSecurity2024}.
However, these artefacts are often incomplete, inconsistently documented, or interpreted differently across roles, leading to gaps in understanding and inconsistent assessments.
Consequently, stakeholders must actively exchange context-specific knowledge to overcome these challenges and manage the complexity of security risk identification.

\paragraph{C3: Continuous System Changes}
The dynamic nature of containerised systems exacerbates these issues.
Frequent updates to container images, deployment settings, or host configurations can invalidate previous risk assessments~\cite{wongSecurityContainersThreat2023, millsLongitudinalRiskbasedSecurity2023}.
Without mechanisms for continuous knowledge exchange and automated reasoning, stakeholders struggle to keep pace with changes, resulting in outdated or incomplete evaluations.

\subsection{Gaps in State-of-the-Art Approaches} \label{sec:state-of-the-art_analysis}
Table~\ref{tab:soa_vs_challenges} summarises our analysis on how state-of-the-art approaches address the identified challenges (\textit{C1--C3}).
We analyse each category and highlight the resulting literature gaps (\textit{Gap1--Gap3}).

\begin{table*}[htbp]
	\caption{Suitability of state-of-the-art approaches with respect to the challenges in security risk identification for container deployments in OT contexts.
		Symbols: \suitable = suitable, \partsuitable = partially suitable, \notsuitable = not suitable.}
	\label{tab:soa_vs_challenges}
	\small%
	\begin{tabularx}{\linewidth}{
			>{\raggedright\arraybackslash}p{0.28\linewidth}
			>{\raggedright\arraybackslash}p{0.21\linewidth}
			>{\raggedright\arraybackslash}p{0.21\linewidth}
			>{\raggedright\arraybackslash}p{0.21\linewidth}}
		\toprule
		\textbf{Approach Category} &
		\textbf{C1: Complexity of Security Risk Identification} &
		\textbf{C2: Need for Context-Specific Knowledge Exchange} &
		\textbf{C3: Continuous System Changes} \\
		\midrule
		
		\textbf{Methods for security risk identification} \newline
		(e.g. CORAS \cite{lundModelDrivenRiskAnalysis2011}, Fault Tree Analysis~\cite{iec61025FaultTree1990}, Attack Trees \cite{schneierAttackTrees1999}, STRIDE \cite{microsoftcorporationSTRIDEThreatModel2009}) &
		\textbf{\suitable} \newline
		\emph{structured joint expert assessment; agreed-upon risk set; well defined processes} &
		\textbf{\partsuitable} \newline
		\emph{knowledge exchange in structured brainstormings; limited generalisation to other contexts} &
		\textbf{\notsuitable} \newline
		\emph{resource-intensive; limited automation; low reproducibility} \\
		
		\textbf{Security controls for containers} \newline
		(e.g. deployment profiles \cite{thekubernetesauthorsPodSecurityStandards2025}, policies \cite{zhuLicSecEnhancedAppArmor2021}, usage control \cite{kelbertFullyDecentralizedData2015}, security assurance \cite{jaskolkaRecommendationsEffectiveSecurity2020}) &
		\textbf{\partsuitable} \newline
		\emph{high maintenance due to complexity; suitable for complex hybrid IT/OT architectures} &
		\textbf{\notsuitable} \newline
		\emph{only general guidance; assessing risks for context-specific artefacts (e.g. Dockerfiles) needs extra effort} &
		\textbf{\suitable} \newline
		\emph{supports continuous review of system changes; deterministic and reproducible once specified} \\
		
		\textbf{Security knowledge management} \newline
		(e.g. vulnerability/threat DBs \cite{werlingerPreparationDetectionAnalysis2010, martinDockerEcosystemVulnerability2018}, cybersecurity knowledge graphs \cite{sikosCybersecurityKnowledgeGraphs2023}, guidelines \cite{souppayaApplicationContainerSecurity2017}, standards \cite{centerforinternetsecurityCISDockerBenchmark2024}) &
		\textbf{\partsuitable} \newline
		\emph{raises general awareness; not context-aware; require additional analysis to assess applicability} &
		\textbf{\suitable} \newline
		\emph{store/share cross-domain knowledge; agreed rulesets; multiple abstraction levels} &
		\textbf{\partsuitable} \newline
		\emph{scans/analyses can be continuous; interpretation is manual; impact depends on knowledge granularity} \\
		\bottomrule
	\end{tabularx}
\end{table*}

\textit{Methods for security risk identification}---such as CORAS~\cite{lundModelDrivenRiskAnalysis2011}, Fault Tree Analysis~\cite{iec61025FaultTree1990}, Attack Trees~\cite{schneierAttackTrees1999}, and STRIDE~\cite{microsoftcorporationSTRIDEThreatModel2009}---support \textit{C1} by providing structured frameworks for joint expert assessment.
They enable stakeholders to co-create threat models and converge on a shared understanding of risks.
However, their contribution to \textit{C2} is limited: knowledge is exchanged primarily during workshops and modelling sessions~\cite{lundModelDrivenRiskAnalysis2011, refsdalCyberRiskManagement2015}, while generalisable security knowledge often remains implicit or undocumented~\cite{lundRiskAnalysisChanging2011}.
For \textit{C3}, these approaches are typically resource-intensive and lack automation~\cite{solhaugModeldrivenRiskAnalysis2014, refsdalSecurityRiskAnalysis2015}.
Their outputs are rarely maintained as machine-readable artefacts, which hinders reproducibility and traceability across evolving OT container deployments.
As a result, assessments are difficult to replicate, and risk identification cannot be consistently integrated into automated workflows.
$\rightarrow$ \textit{\textbf{Gap1}: Methods for security risk identification lack the reproducibility required to address continuous system changes}.

\textit{Security controls for containers}---such as deployment profiles~\cite{thekubernetesauthorsPodSecurityStandards2025}, policy enforcement mechanisms~\cite{zhuLicSecEnhancedAppArmor2021}, usage control~\cite{kelbertFullyDecentralizedData2015}, and assurance practices~\cite{jaskolkaRecommendationsEffectiveSecurity2020}---are well aligned with \textit{C3}, as they support continuous evaluation and deterministic results once specified.
Their support for \textit{C2} is mixed: they incur high maintenance overheads as system complexity grows~\cite{wenDevelopingSecurityAssurance2022, liSecuringServerlessComputing2021}.
However, when sufficiently specified, they can address the complexity of hybrid IT/OT architectures. 
With respect to \textit{C2}, these controls tend to underperform: they provide only general guidance and assessing risks for context-specific artefacts (e.g. Dockerfiles and manifests) requires additional expert effort~\cite{landeckAssuranceApplicationSecurity2024}.
Moreover, they often lack mechanisms to explain why a particular configuration is risky in a given context, making it difficult for non-experts to interpret and act on findings.
$\rightarrow$ \textit{\textbf{Gap2}: Security controls for containers lack the interpretability required to exchange context-specific knowledge}.

\textit{Security knowledge management}---covering vulnerability/threat repositories~\cite{werlingerPreparationDetectionAnalysis2010, martinDockerEcosystemVulnerability2018}, cybersecurity knowledge graphs~\cite{sikosCybersecurityKnowledgeGraphs2023, roweCoordinationModelAttack2023}, formalised rulesets~\cite{doExplainingStaticAnalysis2022, doanDAVSDockerfileAnalysis2022}, guidelines~\cite{souppayaApplicationContainerSecurity2017}, and standards~\cite{centerforinternetsecurityCISDockerBenchmark2024}---best addresses \textit{C2} by enabling storage and sharing of knowledge at multiple abstraction levels.
These resources help raise awareness and provide structured guidance, but their contribution to \textit{C1} is limited: they are often general and not context-aware, requiring additional analysis to assess applicability~\cite{linMeasurementStudyLinux2018, wongSecurityContainersThreat2023}.
For \textit{C3}, scans and analyses backed by these knowledge sources can run continuously, but interpreting results and judging their implications still demands manual effort.
The usefulness of findings depends on the granularity and currency of the underlying knowledge, which must be regularly updated to track the evolving threat landscape~\cite{sroorSystematicMappingStudy2024, alyasContainerPerformanceVulnerability2022}.
Critically, these approaches lack semantic links between risks and the technical artefacts found in container configurations, which limits automation and integration into stakeholder workflows.
$\rightarrow$ \textit{\textbf{Gap3}: Security knowledge management lacks the technical integration required to manage the complexity of security risk identification}.

These gaps motivate our research question:
\begin{quote}\textit{RQ: How can we enable reproducible, interpretable, and technically integrated security risk identification for container deployments in OT contexts?}\end{quote}

\section{Research Design}
We follow a Design Science Research (DSR) approach \cite{peffersDesignScienceResearch2007}, structured around six core phases: problem identification, objectives of a solution, design and development, demonstration, evaluation, and communication.
The main artefact of our research is the \textit{Container Security Risk Ontology (CSRO)}, instantiated as a knowledge graph and supported through a command line tool in our case study.

The research progressed through three iterations, each contributing to the refinement of our approach.
The first iteration (03/2022--03/2023) focused on security assurance for containerised applications, including the analysis of OT requirements and secure container development practices.
The second iteration (01/2023--06/2024) addressed policy-based assessment and management of privileged container deployments in OT contexts, resulting in an earlier version of our tool and templates for policy-compliant container configuration.
The third and current iteration (since 04/2024) targets automated identification of security risks and suitable treatments for OT container deployments.
This paper presents the outcomes of this third iteration, structured according to the DSR phases.

\subsection{Evaluation Case Study}
To evaluate the proposed approach, we conducted a case study of a large-scale industrial software platform for containerised applications in OT contexts.
Our development is informed by the the analysis of the following empirical resources:
\begin{itemize}
	\item > 50 documents on container security in OT contexts, including standards, guidelines, threats, and vulnerabilities,
	\item > 40 discussions with stakeholders---such as security experts, container developers, and system operators---conducted through risk assessments, workshops, and interviews,
	\item > 80 deployment configuration files of containerised applications deployed on the platform.
\end{itemize}

\subsection{Ontology Engineering}
The design of \textit{CSRO} follows established ontology engineering principles~\cite{noyGuideCreatingYour2001, suarez-figueroaNeOnMethodologyOntology2012}, including:
\begin{itemize}
	\item reuse of existing domain vocabularies where applicable,
	\item separation of conceptual and implementation layers,
	\item iterative refinement based on stakeholder feedback,
	\item alignment with OWL~2 and RDF standards for semantic interoperability.
\end{itemize}

Ontology engineering is inherently iterative~\cite{poveda-villalonLOTIndustrialOriented2022}, making it well suited to our DSR approach.
During the first and second iterations of our research, we captured relevant security knowledge on requirements, practices, and policy-based assessment in an ontology for security assurance of containerised applications.
This knowledge is not explicitly presented in this paper, but presents opportunities to further enhance \textit{CSRO}.

\section{Objectives of a Solution} \label{sec:objectives}
The goal of our research is to design a solution that addresses the literature gaps in security risk identification for OT container deployments (cf. \textit{Gap1--Gap3}, Section~\ref{sec:state-of-the-art_analysis}).
In this section, we present four objectives of a solution (\textit{O1--O4}) to achieve this goal.

\paragraph{O1: Reproducible Security Risk Identification}
We aim to provide an approach that makes the security risk identification process more reproducible.
To align with the continuous change management of OT container deployments, the process must be deterministic---i.e. returning the same risks for the same input.
This objective reflects ongoing efforts in artificial intelligence research to implement consistency and transparency in context-dependent processes~\cite{gundersenStateArtReproducibility2018, kimGenerativeArtificialIntelligence2024}.
In risk assessment research, reproducibility is also pursued to meet operational and economic requirements~\cite{haasReproducibleRiskAssessment2016, carvalhoStabilityReproducibilitySemiquantitative2015}.
By formalising the risk identification process and enabling automated reasoning, we aim to complement existing methods and address \textit{Gap~1}.

\paragraph{O2: Completeness of Identified Security Risks}
We aim to ensure that our approach identifies all relevant security risks for OT container deployments.
Completeness is a key quality criterion in risk assessment, and methods such as CORAS are considered complete when structured brainstorming is used to exhaustively review threats~\cite{lundModelDrivenRiskAnalysis2011}.
Existing work has proposed frameworks to compare the completeness of different methods~\cite{wangenFrameworkEstimatingInformation2018, ghazouaniInformationSecurityRisk2014}.
Our solution should be capable of identifying risks covered by state-of-the-art approaches, and additionally support the discovery of risks that emerge from combinations of container traits or context-specific configurations.
This objective further supports the reproducibility of security risk identification highlighted in \textit{Gap~1}.

\paragraph{O3: Multifaceted Risk Representations}
Our approach should support multifaceted representations of risks to improve interpretability and stakeholder relevance.
A facet represents a “hierarchy of homogeneous terms describing an aspect of the domain”~\cite{giunchigliaFacetBasedMethodologyConstruction2012}.
Multifaceted risk representations allow complex threats, impacts, and treatments to be tailored to the perspective of different stakeholders.
Techniques include categorising risks, visualising likelihood and risk calculation (e.g. risk matrices), and linking risks to specific treatment strategies (e.g. configuration hardening)~\cite{menoniAssessingMultifacetedVulnerability2012, robertsonMultifacetedHolisticRisk2022}.
This objective supports interpretability across contexts (\textit{Gap~2}) and introduces semantic relations between risks and technical artefacts to enhance integration (\textit{Gap~3}).

\paragraph{O4: Enabling Pervasive Technical Integration}
We aim to enable pervasive technical integration of our solution---i.e. seamless interaction with diverse technologies, applications, data sources, and devices~\cite{liuPervasiveInformaticsTheory2010, cantnerPervasiveTechnologiesIndustrial2021}.
Security risk identification should not be treated as a siloed or post-hoc activity, but as a continuous and context-aware process embedded into the lifecycle of container deployments.
Inspired by digital platform engineering~\cite{gawerDigitalPlatformsEcosystems2022}, our approach should support end-to-end integration of risks and treatments into stakeholder workflows.
This objective contributes to all identified gaps: enabling reproducibility (\textit{Gap~1}), supporting interpretability (\textit{Gap~2}), and fostering technical integration (\textit{Gap~3}).

\section{Design and Development}
We propose a model-based approach for security risk identification of OT container deployments.
In this context, a model refers to an abstract, structured representation of relevant system elements~\cite{bkcaseeditorialboardGuideSystemsEngineering2025}.
These models serve as shared reference points for stakeholders, fostering a common understanding and supporting the identification of context-specific security risks.

Our approach incorporates knowledge representation to encode domain knowledge in a formal structure that is both machine-readable and human-interpretable~\cite{brachmanKnowledgeRepresentationReasoning2004}.
This enables integration of diverse information sources and supports automated analysis.
We also apply reasoning mechanisms---formal semantics and logical inference---to derive insights from existing knowledge~\cite{krotzschDescriptionLogics2013}.
These mechanisms allow stakeholders to infer threats, validate assumptions, and assess the impact of changes, supporting a proactive and adaptive risk identification process.

\begin{figure}[htbp]
	\centering
	\includegraphics[width=\linewidth]{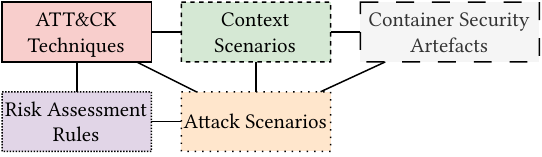}
	\caption{Overview of the five domains of our \textit{Container Security Risk Ontology (CSRO)}}
	\label{fig:ontology_concepts_simple}
	\Description{The figure provides a simplified overview of the five domains in the Container Security Risk Ontology (CSRO). It includes ATT\&CK Techniques, Context Scenarios, Attack Scenarios, Risk Assessment Rules, and Container Security Artefacts. These domains represent the core structure of the ontology used for identifying and assessing security risks in container deployments.}
\end{figure}

In the following, we present the design of our knowledge model, which is implemented as the \textit{Container Security Risk Ontology (CSRO)}.
\textit{CSRO} integrates five key domains:
\textit{ATT\&CK Techniques}, \textit{Context Scenarios}, \textit{Attack Scenarios}, \textit{Risk Assessment Rules}, and \textit{Container Security Artefects} (cf. Figure~\ref{fig:ontology_concepts_simple}).

\subsection{ATT\&CK Techniques Targeting Containers}
Threat modelling is a well-established practice in cybersecurity to identify and mitigate threats before exploitation~\cite{xiongThreatModelingSystematic2019}.
MITRE ATT\&CK is a curated knowledge base of adversary tactics and techniques based on real-world observations~\cite{stromMITREATTCKDesign2018}, including a dedicated view on container-targeted attacks\footnote{cf. \url{https://attack.mitre.org/versions/v17/matrices/enterprise/containers/}}.

We integrate the terminology of MITRE ATT\&CK directly into \textit{CSRO}, focusing on the concept of an \textit{ATT\&CK Technique}, which describes how an adversary achieves a tactical objective.
The concept of a Sub-technique is excluded for simplicity, as it shares the same modelling characteristics.

For example, the shellcode injection attack introduced in Section~\ref{sec:security_risk_assessment} corresponds to the \textit{Process Injection: Ptrace System Calls} \textit{Technique}\footnote{cf. \url{https://attack.mitre.org/versions/v17/techniques/T1055/008/}}.
This integration allows us to reference ATT\&CK resources such as procedure examples and mitigations, and benefit from its continuous evolution.

\subsection{Context Scenarios Based on Container Security Assumptions} \label{sec:context_scenarios_security_assumptions}
Context establishment is a preparatory activity in risk assessment~\cite{internationalorganizationforstandardizationISO31000Risk2018}.
It involves defining the target of assessment and the assumptions about its environment~\cite{refsdalCyberRiskManagement2015}.

In our case, the target is a containerised application in an OT context.
Security assumptions---such as whether a container is exposed to external services or restricted by network zones---greatly influence the associated risks.
Other assumptions concern container properties like software footprint or runtime privileges.

To represent these aspects, \textit{CSRO} includes \textit{Container Security Assumptions}.
We aggregate related assumptions into \textit{Context Scenarios}, which define the baseline for assessing risks associated with an \textit{ATT\&CK Technique}.
This design is inspired by scenario-based techniques used in domains such as automated driving~\cite{menzelScenariosDevelopmentTest2018, kolbWhenApplyScenarioBased2023}, and adapted for risk evaluation~\cite{wangScenariobasedRiskEvaluation2021, khatunApproachScenarioBasedThreat2021}.

\subsection{Attack Scenarios} \label{sec:attack_scenarios}
To link \textit{ATT\&CK Techniques} and \textit{Context Scenarios}, we introduce the domain of \textit{Attack Scenarios}.
This represents a context-specific threat instance and is inspired by model-based risk assessment methods such as CORAS~\cite{lundModelDrivenRiskAnalysis2011}.
These methods aim to improve stakeholder understanding by modelling threats, impacts, and consequences, and by supporting the identification of suitable risk treatments~\cite{solhaugCORASLanguageWhy2014}.

\subsection{Risk Assessment Rules}
In \textit{CSRO}, we adapt \textit{Risk Assessment Rules} that define a risk level as the combination of likelihood and impact of an attack.
Impact is determined based on the potential damage to valued security assets, such as the availability of an industrial control system.
Likelihood, on the other hand, is derived by evaluating the exposure of a system’s weakness and the exploitability of the attack technique.
These factors are influenced by the satisfaction of security assumptions defined during risk assessment.
We observe this approach in our case study and recognise similar practices in the literature~\cite{freundMeasuringManagingInformation2014, refsdalCyberRiskManagement2015}.

\subsection{Container Security Artefacts}
Finally, our knowledge model includes \textit{Container Security Artefacts}, which link technical artefacts to the other domains of \textit{CSRO}.
These artefacts include deployment configurations, vulnerability scan results, container security guidelines, and standards---such as the NIST Application Container Security Guide~\cite{souppayaApplicationContainerSecurity2017}.

In \textit{CSRO}, we model container deployment settings---such as \texttt{pid=host} and \texttt{cap-add=SYS\_PTRACE}---as individual artefacts.
We define semantic links from combinations of these artefacts to specific \textit{ATT\&CK Techniques}, enabling automated assessment of whether the technical requirements for an attack are met in a given container context.
Container security standards are also represented as artefacts, allowing us to link deployment settings to relevant sections of these standards to support compliance evaluation.
Furthermore, we associate artefacts with security assumptions, enabling automated derivation of their satisfaction state based on configuration files or vulnerability scan results.
This integration supports context-aware risk identification by assessing the relevance of context scenarios through artefact-based reasoning.

\section{Demonstration}
\subsection{Container Security Risk Ontology (\textit{CSRO})} \label{sec:ontology_concepts}
\begin{figure*}[htbp]
	\centering
	\includegraphics[width=\textwidth]{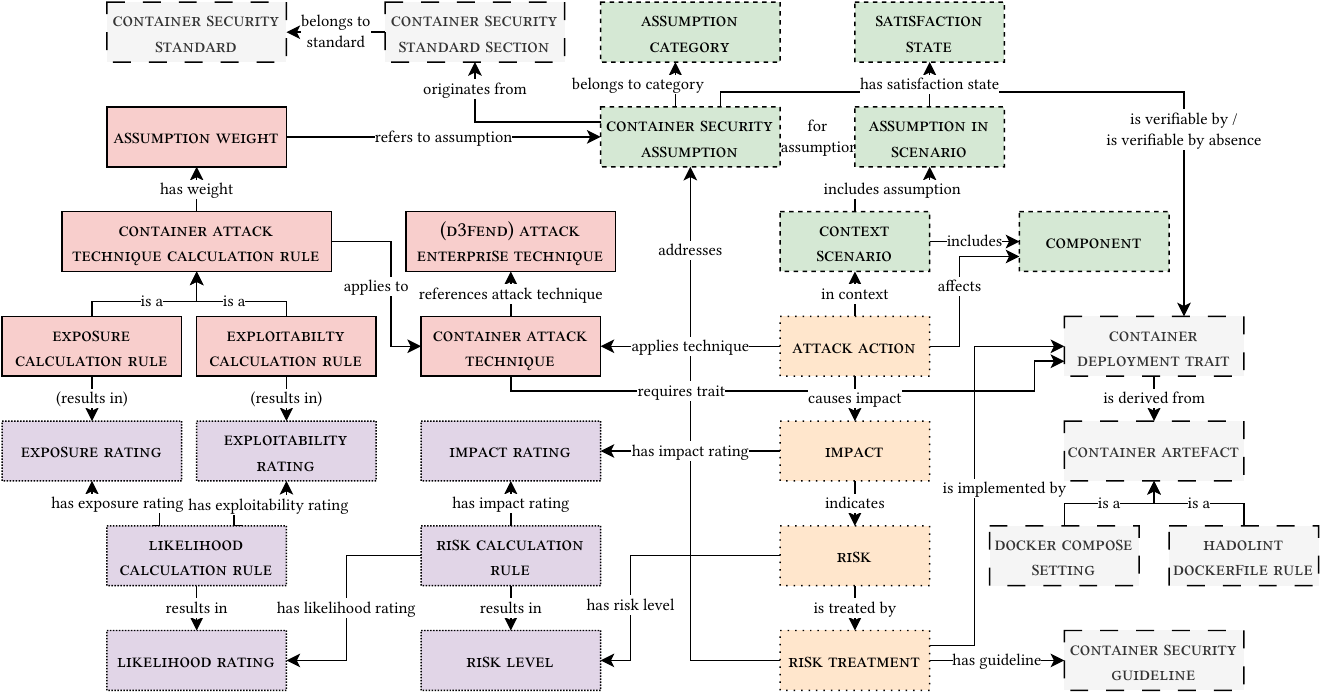}
	\caption{Core concepts and relationships in the \textit{Container Security Risk Ontology (CSRO)}.
		The ontology integrates five domains: ATT\&CK Techniques (red, solid), Context Scenarios (green, dashed), Attack Scenarios (orange, wide-dotted), Risk Assessment Rules (purple, dotted), and Container Security Artefacts (grey, wide-dashed).
		Prefixes and namespace annotations are omitted.}
	\label{fig:ontology_concepts}
	\Description{The figure presents the core concepts and relationships in the Container Security Risk Ontology (CSRO). It includes five domains: ATT\&CK Techniques, Context Scenarios, Attack Scenarios, Risk Assessment Rules, and Container Security Artefacts. The diagram shows how container traits, assumptions, and artefacts are linked to attack techniques and risk treatments. It also illustrates how exploitability and exposure ratings are calculated from weighted assumptions, combined into likelihood ratings, and used to derive overall risk levels. References to standards, guidelines, and tools such as Hadolint and Docker Compose are included to support technical integration.}
\end{figure*}

This section demonstrates the practical application of our approach by presenting the \textit{Container Security Risk Ontology (CSRO)}.\footnote{The ontology, the full list of container security assumptions, and risk assessment rules are available at \url{https://w3id.org/csro}}
\textit{CSRO} formalises the knowledge model introduced in the previous section and serves as the foundation for automated risk identification for containers in OT contexts.
Figure~\ref{fig:ontology_concepts} provides an overview of the core concepts and their relationships within \textit{CSRO}.

Within the domain of \textit{ATT\&CK Techniques}, we define the class \textsc{Container Attack Technique}, which includes one or more required \textsc{Container Deployment Trait}, used to verify whether a container enables a specific attack.
Each technique is linked to a \textsc{Calculation Rule}, including \textsc{Exploitability Calculation Rule} and \textsc{Exposure Calculation Rule}, which use weighted \textsc{Container Security Assumption} inputs to estimate likelihood.
The MITRE ATT\&CK framework is not directly provided as an ontology or knowledge graph.
To enable semantic integration, we reference the MITRE D3FEND ontology\footnote{cf. \url{https://d3fend.mitre.org}}, which is a knowledge graph of cybersecurity countermeasures also developed and maintained by MITRE.
D3FEND includes stable references to all ATT\&CK Tactics and Techniques, making it a valuable resource for linking our ontology to ATT\&CK concepts.
We use the concept \textsc{AttackEnterpriseTechnique} from D3FEND to establish these links.

In the domain of \textit{Context Scenarios}, we represent \textsc{Context Scenario} as an aggregation of \textsc{Component} and associated \textsc{Assumption in Scenario} with corresponding \textsc{Satisfaction State}.
Each assumption belongs to an \textsc{Assumption Category} and originates from a \textsc{Container Security Standard Section}, which references a specific section in a \textsc{Container Security Standard}, such as §4.1.2 in the NIST Application Container Security Guide~\cite{souppayaApplicationContainerSecurity2017} or §5.24 of the CIS Docker Benchmark~\cite{centerforinternetsecurityCISDockerBenchmark2024}.
Table~\ref{tab:security_assumptions} presents a selection of instantiated container security assumptions.
The set of \textsc{Container Security Assumption} can be extended beyond container traits to include assumptions about the host system or the OT environment.

\begin{table}[htbp]
	\caption{Excerpt of the container security assumptions in \textit{CSRO}, curated from container security standards.}
	\label{tab:security_assumptions}
	\begin{tabular}{@{}p{0.6cm}p{1.2cm}p{4.0cm}p{1.6cm}@{}}
		\toprule
		\textbf{ID} & \textbf{Category} & \textbf{Assumption Description} & \textbf{Source} \\
		\midrule
		RTS\_1 & Runtime & App containers are run as non-root user & §4.1.2 NIST, §5.24 CIS, ... \\
		RTS\_2 & Runtime & App containers use read-only file systems & §4.4.4 NIST, §5.13 CIS, ... \\
		RTS\_3 & Runtime & App containers are deployed with minimal Linux capabilities & §4.4.3 NIST, §4.8 CIS, ... \\
		IMG\_1 & Image & App images are scanned for vulnerabilities & §4.1.1 NIST, §4.4 CIS, ... \\
		NET\_1 & Network & App container networks are segmented & §4.3.3 NIST, §5.10 CIS, ... \\
		NET\_2 & Network & Host-level firewall rules and network policies are applied & §7.3 CIS, ... \\
		... & ... & ...& ... \\
		\bottomrule
	\end{tabular}
\end{table}

The domain of \textit{Attack Scenarios} introduces the class \textsc{Attack Action}, which instantiates a \textsc{Container Attack Technique} in a specific \textsc{Context Scenario}.
Each action causes one or more \textsc{Impact}, assigned an \textsc{Impact Rating}, and indicates a corresponding \textsc{Risk}.
Risks can be mitigated by a \textsc{Risk Treatment}, which addresses specific \textsc{Container Security Assumption}, references a \textsc{Container Security Guideline}, and is implemented via a \textsc{Container Deployment Trait}.

In the domain of \textit{Risk Assessment Rules}, we define \textsc{Exposure Rating} and \textsc{Exploitability Rating}, which are combined by a \textsc{Likelihood Calculation Rule} to derive a \textsc{Likelihood Rating}.
A \textsc{Risk Calculation Rule} then derives a \textsc{Risk Level} from \textsc{Likelihood Rating} and \textsc{Impact Rating}.

Finally, the domain of \textit{Container Security Artefacts} includes classes such as \textsc{Container Security Standard}, \textsc{Container Deployment Trait}, and \textsc{Container Security Guideline}.
These link the ontology to real-world \textsc{Container Artefact} such as \textsc{Docker Compose Setting} or \textsc{Hadolint Dockerfile Rule}, enabling integration with tools for automated risk identification.

\begin{figure}[htbp]
	\centering
	\includegraphics[width=\linewidth]{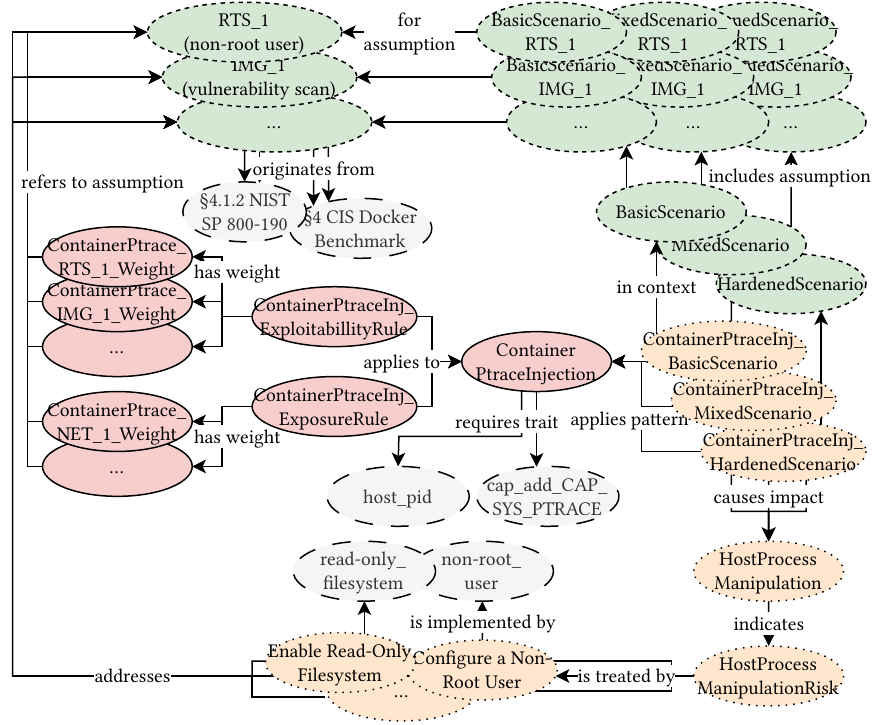}
	\caption{\textit{CSRO} knowledge graph instances of the shellcode injection technique for three context scenarios.}
	\label{fig:ontology_instances}
	\Description{The figure shows instantiated individuals of the Container Security Risk Ontology (CSRO) for the attack technique ContainerPtraceInjection across three context scenarios: BasicScenario, MixedScenario, and HardenedScenario. Each scenario includes assumptions such as IMG\_1 (vulnerability scan) and RTS\_1 (non-root user), with associated weights used in exploitability and exposure calculation rules. The attack causes the impact HostProcessManipulation, which is linked to a risk and treated by configuration traits like non-root user and read-only filesystem. References to container security standards such as NIST SP 800-190 and the CIS Docker Benchmark are included to support compliance.}
\end{figure}

\begin{framed}
	\noindent
	\textbf{Running example: instantiation of the shellcode injection technique for three context scenarios (Figure \ref{fig:ontology_instances}).}
	
	We instantiate the \textsc{Container Attack Technique} \textit{ContainerPtraceInjection}, which requires the traits \textit{host\_pid} and \textit{cap\_add\_CAP\_SYS\_PTRACE}.
	These traits indicate that the container shares the host’s process namespace and has permission to trace other processes, enabling shellcode injection.
	
	To estimate the likelihood of this attack, we define two \textsc{Calculation Rule} individuals: \textit{ContainerPtraceInjectionExploitabilityRule} and \textit{ContainerPtraceInjectionExposureRule}.
	Each rule assigns weights to relevant \textsc{Container Security Assumption} individuals, such as \textit{RTS\_1} (non-root user), \textit{IMG\_1} (vulnerability scan), and \textit{NET\_1} (network segmentation).
	
	For this example, we instantiate three \textsc{Context Scenario} individuals: \textit{BasicScenario}, \textit{MixedScenario}, and \textit{HardenedScenario}.
	Each scenario includes a set of \textsc{Assumption in Scenario} individuals with associated \textsc{Satisfaction State}, e.g. \textit{RTS\_1} is satisfied.	
	Next, we define three \textsc{Attack Action} individuals, each linking the technique to one of the context scenarios.
	Each action causes the same \textsc{Impact} (\textit{HostProcessManipulation}) and indicates the corresponding \textsc{Risk} (\textit{HostProcessManipulationRisk}).
	
	We also instantiate \textsc{Risk Treatment} individuals, such as configuring a non-root user or enabling a read-only filesystem.
	These treatments are linked to the assumptions they address and the traits used to implement them.
	Finally, we reference related \textsc{Container Security Standard Section} and associated \textsc{Container Deployment Trait} to support compliance evaluation and automated reasoning.
\end{framed}

\subsection{Risk Level Calculation with SPARQL} \label{sec:risk_level_calculation}
To calculate the risk level of each attack, we query the instantiated \textit{CSRO} knowledge graph using SPARQL.\footnote{SPARQL is a W3C-standardised query language for RDF data. See \url{https://www.w3.org/TR/sparql11-query/}. Our queries are available at \url{https://w3id.org/csro}}
The process evaluates each \textsc{Attack Action} and derives ratings based on contextual assumptions and predefined rules.
We adapt matrix-based calculations for likelihood and risk level from established literature~\cite{freundMeasuringManagingInformation2014, refsdalCyberRiskManagement2015}, but introduce our own formal rules for computing exploitability and exposure, which are not specified in prior work.
To simplify modelling and ensure maintainability, we statically assign \textsc{Impact Ratings} to an \textsc{Attack Action}; dynamic impact calculation would require stakeholder-specific protection goals and introduce significant complexity, which we leave for future work.
Algorithm~\ref{alg:risk_calculation} outlines the procedure for calculating risk ratings and treatments for all instances of \textsc{Attack Action}.

\begin{algorithm}[htbp]
	\caption{Calculate Risk Ratings and Treatments.}
	\label{alg:risk_calculation}
	\SetKwInOut{Input}{Input}
	\SetKwInOut{Output}{Output}
	\LinesNumbered
	\DontPrintSemicolon
	\Input{\textit{CSRO Knowledge Graph}}
	\Output{Security risks for a container with calculated ratings and treatments}
	\ForEach{\textsc{Attack Action} in \textit{CSRO} Knowledge Graph}{
		Retrieve associated \textsc{Container Attack Technique}, \textsc{Context Scenario}, and affected \textsc{Component}\;
		Extract \textsc{Container Security Assumption} states from \textsc{Assumption in Scenario}\;
		Compute \textsc{Exploitability Rating} and \textsc{Exposure Rating} using rule-based scoring and thresholds\;
		Derive \textit{Likelihood} and \textit{Risk Level} from ratings\; 
		Retrieve applicable \textsc{Risk Treatment} with guidelines\;
		\textbf{CONSTRUCT} OWL triples for risks, ratings, and treatments\;
	}
\end{algorithm}

\paragraph{Computation of \textsc{Exploitability Rating} and \textsc{Exposure Rating}}
To assess the exploitability and exposure of an \textsc{Attack Action}, our model combines the satisfaction states of \textsc{Container Security Assumption} individuals in a given \textsc{Context Scenario} with rule-specific weights defined by the associated \textsc{Calculation Rule}.
Each \textsc{Container Security Assumption} $i$ has a satisfaction score $s_i$ based on its fulfilment in the scenario:

\begin{equation}
	s_i =
	\begin{cases}
		1.0 & \text{if Satisfied} \\
		0.5 & \text{if Unknown} \\
		0.0 & \text{if Dissatisfied}
	\end{cases}
\end{equation}

Each \textsc{Container Attack Technique} defines a set of weights $w_i$ for relevant assumptions via its \textsc{Exploitability Calculation Rule} and \textsc{Exposure Calculation Rule}.
These weights indicate how effective a satisfied assumption is in reducing the exploitability or exposure of the technique.

To compute the security score $\Sexp$ for exploitability, we multiply each assumption's weight $w_i$ with its satisfaction score $s_i$ and sum the results:

\begin{equation}
	\Sexp = \sum_{i \in \Aex} s_i \cdot w_i
\end{equation}

We then define dynamic weight thresholds $\theta$ to map the score to a rating.
The total weight $\Wex$ for all assumptions in the rule is:

\begin{equation}
	\Wex = \sum_{i \in \Aex} w_i
\end{equation}

The thresholds divide the score range $[0, \Wex]$ into three equal intervals:

\begin{align}
	\thi{\mathrm{exp}} &\coloneqq \tfrac{1}{3} \Wex \\
	\tlow{\mathrm{exp}}  &\coloneqq \tfrac{2}{3} \Wex
\end{align}

The exploitability rating is derived as follows:

\begin{equation}
	\text{Exploitability}(\Sexp) =
	\begin{cases}
		\textit{Low}    & \text{if } \Sexp \ge \tlow{\mathrm{exp}} \\
		\textit{Medium} & \text{if } \tlow{\mathrm{exp}} \le \Sexp < \thi{\mathrm{exp}} \\
		\textit{High}   & \text{if } \Sexp < \thi{\mathrm{exp}}
	\end{cases}
\end{equation}

The same procedure is applied to compute the exposure rating, using the set $\Aexpos$ and corresponding thresholds $\thi{\mathrm{expo}}, \tlow{\mathrm{expo}}$.

\medskip

\noindent
\emph{Note.} The computed score $S$ reflects the effectiveness of mitigations in the scenario.
Higher scores indicate a stronger security posture, which results in a lower exploitability or exposure rating.
This reflects that both exploitability and exposure are risk factors that should be minimised.
Also, this scoring approach represents a simplified heuristic for demonstration; in practice, more complex rules may be required to accurately estimate exploitability and exposure.

\begin{framed}
	\noindent
	\textbf{Running example: exploitability and exposure calculation for \textit{ContainerPtraceInjection} in \textit{MixedScenario}.}
	
	\noindent
	In \textit{MixedScenario}, the following \textsc{Container Security Assumption} states apply:
	\begin{itemize}
		\item \textit{RTS\_1}, \textit{RTS\_3}: Satisfied
		\item \textit{RTS\_2}, \textit{IMG\_1}, \textit{NET\_1}: Dissatisfied
		\item \textit{NET\_2}: Unknown
	\end{itemize}
	
	The \textit{ExploitabilityRule} defines the weights: $w(\textit{RTS\_1})=3$, $w(\textit{RTS\_2})=2$, $w(\textit{RTS\_3})=1$, $w(\textit{IMG\_2})=1$.
	The resulting total weight is: $\Wex = 7$.
	Resulting in the thresholds: $\thi{\mathrm{exp}} = \tfrac{7}{3} \approx 2.33$, $\tlow{\mathrm{exp}} = \tfrac{14}{3} \approx 4.67$.
	Therefore, $\Sexp = 3 + 0 + 1 + 0 = 4 \Rightarrow$ \textsc{Exploitability Rating}: \textit{Medium}
	
	The \textit{ExploitabilityRule} defines the weights:: $w(\textit{NET\_1})=2$, $w(\textit{NET\_2})=2$.
	The resulting total weight is: $\Wexpos = 4$. 
	Resulting in the thresholds: $\thi{\mathrm{expo}} = \tfrac{4}{3} \approx 1.33$, $\tlow{\mathrm{expo}} = \tfrac{8}{3} \approx 2.67$.
	Therefore, $\Sexpos = 0 + 1 = 1 \Rightarrow$ \textsc{Exposure Rating}: \textit{High}
	
	These ratings result in the \textsc{Likelihood Rating} \textit{Likely}.  
	With an \textsc{Impact Rating} of \textit{Disastrous} for \textit{HostProcessManipulation}, the resulting \textsc{Risk Level} is \textit{Major} (rule matrices are omitted for simplicity, cf. \url{https://w3id.org/csro} for details).
\end{framed}

\subsection{Integration as a Command Line Tool}
We integrate \textit{CSRO} into stakeholder workflows using a command-line tool that parses container artefacts and generates a structured risk factsheet.
The factsheet is based on the output of the risk calculation Algorithm~\ref{alg:risk_calculation}, which derives ratings and treatments for instantiated attack actions.
The tool currently analyses:
\begin{itemize}
	\item \texttt{docker-compose.yml} files to extract deployment traits,
	\item \texttt{Dockerfile} content using the Hadolint linter~\cite{hadolintcontributorsGitHubHadolintHadolint2022} to identify vulnerabilities.
\end{itemize}

To evaluate \textsc{Container Security Assumption} satisfaction, the tool compares extracted traits against verification rules defined in the \textit{CSRO} knowledge graph.
Each assumption may be verifiable by the presence or absence of one or more traits.
The tool assigns a satisfaction state to each assumption and includes this information in the generated factsheet.

To select the most appropriate \textsc{Context Scenario}, the tool compares the observed trait set against all predefined scenarios.
It computes a fit score based on how many assumptions match and prioritises scenarios with more perfect matches.
This ensures that risk identification remains context-aware and tailored to the container's actual configuration.

The tool then identifies applicable \textsc{Attack Action} individuals by checking whether the required traits for the \textsc{Container Attack Technique} are present.
For each match, it retrieves the associated \textsc{Impact}, \textsc{Exploitability Rating}, \textsc{Exposure Rating}, \textsc{Likelihood Rating} and \textsc{Risk Level}, and adds all information to the factsheet.

Finally, the tool includes suitable \textsc{Risk Treatment} options and referenced \textsc{Container Security Guideline} resources.
This enables tailored guidance for risk mitigation, allowing users to understand the effectiveness of a treatment based on the associated assumption's weight in relation to the attack technique.

\section{Evaluation}
\subsection{Fulfilment of the Objectives}
\paragraph{O1: Reproducible Security Risk Identification}
Our approach enables a formal, reproducible process for identifying security risks in OT container deployments.
By instantiating \textit{CSRO} as a semantic knowledge graph, we provide a structured, machine-interpretable representation of risks associated with attack techniques, tailored to specific operational contexts.
Relations between container security artefacts and risk model entities allow our tool to automatically derive container traits from technical artefacts---such as Docker Compose settings or Hadolint scan results---and verify the satisfaction of security assumptions.
We define deterministic calculation rules for \textit{ATT\&CK Techniques} in container contexts, estimating attack likelihood based on exploitability and exposure, and derive risk levels through transparent, rule-based mappings.
This end-to-end formalisation---from artefact to risk level---enables traceable and automatable risk identification, directly supporting reproducibility.
In our case study, stakeholders appreciated the formalisation and automation, noting that a deterministic baseline improves efficiency across the application lifecycle.
They emphasized that automation should complement, not replace, traditional threat and risk assessment workshops, which remain essential for addressing edge cases and context-specific risks.

\paragraph{O2: Completeness of Identified Security Risks}
While no security risk knowledge base can guarantee absolute completeness, we evaluated \textit{CSRO}’s coverage against the security management practices of the case study platform.
The platform enforces security policies during application publishing, primarily using a blacklist approach to flag critical deployment settings (e.g. \texttt{pid=host}).
We ensured that all policy checks were reflected as an instantiated \textsc{Attack Scenario} in the knowledge graph, achieving completeness with respect to existing policies.
Additionally, \textit{CSRO} captures risks not addressed by current policies, including combinations of traits that are benign in isolation but critical in combination.
Together with domain experts, we systematically reviewed all Docker Compose settings and their combinations for security relevance.
For each relevant setting, we instantiated the associated container attack techniques, impacts, and calculation rules, and validated them with stakeholders.
Maintaining completeness requires ongoing curation of the calculation rules that map techniques to assumptions---a key area for future improvement.

\paragraph{O3: Multifaceted Risk Representations}
The case study platform involves stakeholders with diverse backgrounds, including developers unfamiliar with container security and operators responsible for enforcing security policies.
This diversity creates challenges in translating operational requirements into effective technical treatments.
Our approach supports multifaceted risk representations that bridge these gaps.
Developers can use the command line tool to analyse their applications for known threats, review identified risks, and receive tailored treatment suggestions---including guidelines and templates.
Operators can prioritise risks based on OT context, assess container artefacts (e.g. Docker Compose settings), and evaluate compliance with security standards via linked assumptions.
The modular structure of \textit{CSRO} supports multiple technologies (documents, configurations, rulesets) and varying levels of abstraction.
For example, it can represent the general risk of containers with the \texttt{privileged} flag or model specific attacks like shellcode injection enabled by multiple traits.
\textit{CSRO} is also extensible to broader OT contexts, such as host firewall configurations.
This flexibility makes our approach well-suited to the multifaceted nature of security risk identification in OT container deployments.

\paragraph{O4: Enabling Pervasive Technical Integration}
Instantiating \textit{CSRO} as a semantic graph requires initial knowledge engineering effort.
We mitigate this by reusing established resources such as MITRE ATT\&CK and container security standards.
However, risk calculation rules still require careful instantiation and ongoing maintenance.
The OWL~2-based knowledge graph supports flexible technical interfaces.
For example, the case study platform uses Open Policy Agent (OPA) for policy enforcement. 
\textit{CSRO} allows to both query and extend the knowledge base---e.g. by instantiating new traits based on OPA policies.
SPARQL templates can define technical interfaces, enabling stakeholders to interact with the knowledge base without deep expertise in knowledge engineering.
This supports pervasive integration by embedding risk identification and treatment into existing workflows and technologies.

\subsection{Limitations}
Our current approach focuses on container deployment traits from Docker Compose and Hadolint, but its modular design allows extension to other technologies such as Podman.
Still, we can currently only verify security assumptions that relate directly to the container itself.
Security assumptions concerning the host system and operational technology (OT) controls---such as network segmentation or access restrictions to host devices---are not yet supported by our knowledge graph or tool integration.
To address these limitations, we currently offer a simplified mechanism: users can load an ``assumptions config'' file to manually confirm the satisfaction state of selected assumptions.
For example, users may indicate that host-level firewall rules are applied, satisfying assumption \textit{NET\_2}.
Making such assumptions during risk identification is common practice, as every assessment operates within a defined technical scope.
Our implementation serves as an intermediate solution until more comprehensive technological support is integrated, such as parsing host firewall configuration files.

\textit{CSRO} currently supports only technological treatments for risks, such as hardening container configurations or modifying containerised application architectures.
However, effective risk treatment also includes organisational measures, such as managing access permissions or conducting security training to raise awareness of threats.
In some cases, the outcome of a risk assessment may be to accept a risk.
This occurs when treatments are infeasible due to technical constraints, cost, or potential interference with the container's intended functionality.
In future work, we plan to extend our model to represent the implications of organisational treatments on a pervasive technical basis.
For example, our approach could be used to generate risk assessment reports tailored to specific stakeholder interfaces.
In the case study, Excel questionnaires are used to document the security properties of containerised applications.
These could be auto-filled using our ontology-based approach to improve consistency and reduce manual effort.

Creating and maintaining the \textit{CSRO} knowledge graph requires substantial knowledge engineering effort.
The practical usability of our approach depends on its maintainability over time.
If instantiated risks become outdated and are not updated, the value of the generated assessment results diminishes significantly.
It is therefore essential to reuse existing security knowledge and minimise the overhead of maintaining \textit{CSRO}.
We partially achieve this by referencing the MITRE ATT\&CK knowledge base and established container security guidelines.
However, the calculation rules for exploitability and exposure still require ongoing maintenance and introduce necessary overhead.
To reduce this complexity, we plan to cluster attack techniques by strategy---such as those defined by MITRE ATT\&CK Tactics---and map the effectiveness of security assumptions to these clusters directly.
This should be feasible, as an evolving rule set will likely reveal that certain security controls---such as network segmentation---are broadly effective against a group of related attacks, such as lateral movement.

\subsection{On the Integration of Generative AI}
Generative AI (GenAI) provides an efficient solution for risk identification based on its extensive knowledge of container security~\cite{shuklaGenAICode2023, sroorManagingSecurityIssues2025, zemichealLLMAgentsVulnerability2024}.
It is therefore a valid question whether our approach justifies the modelling effort compared to risk identification using GenAI.
We argue that GenAI is not (yet) suitable to fulfil all desired objectives of a solution.
Its non-deterministic nature~\cite{ouyangEmpiricalStudyNondeterminism2025} fails to address \textit{O1}, as it cannot guarantee a reproducible approach to security risk identification.
Furthermore, the completeness of identified risks (\textit{O2}) and associated treatments can only be verified with the support of security experts, which again introduces additional overhead.
Nevertheless, GenAI excels at addressing objectives \textit{O3} and \textit{O4}, as it can tailor its output to the needs of the user and adapt to different stakeholder perspectives~\cite{shiHCIcentricSurveyTaxonomy2023}.
Ongoing developments towards agentic AI will further improve the pervasiveness of technical integration, for example, by generating hardened container configurations or removing vulnerable software dependencies directly for the user \cite{zemichealLLMAgentsVulnerability2024}.
These findings motivate us to investigate the integration of GenAI alongside our approach.
In the case study, we analysed the potential of using GenAI to further integrate the results of our tool---the risk factsheet---into stakeholder workflows.
The models produced strong results when tasked with summarising the risks of the JSON-based factsheet or guiding specific treatments for the user's context.

\section{Related Work}
\paragraph{Risk Identification Methods}
The CORAS method~\cite{lundModelDrivenRiskAnalysis2011} provides a model-based approach for threat modelling and risk assessment.
It relies on manual identification, often through structured brainstorming sessions.
While extensions have addressed evolving systems~\cite{lundRiskAnalysisChanging2011, refsdalSecurityRiskAnalysis2015}, emerging risks still require manual analysis.

Fault Tree Analysis (FTA)~\cite{iec61025FaultTree1990}, traditionally used in safety and reliability engineering, has also been applied to containerised systems~\cite{bakhshiFaulttolerantPermanentStorage2021, zangFaultTreeBased2019}. 
FTA is effective for modelling failure propagation, but container security often demands a stronger focus on adversarial behaviour and threat modelling.

Attack Trees~\cite{schneierAttackTrees1999} offer a structured way to model threats and their logical relationships.
They are widely used in container security research~\cite{vsContainerSecurityPrecaution2023, sultanContainerSecurityIssues2019, linMeasurementStudyLinux2018}, but their complexity can hinder interdisciplinary collaboration and scalability.

STRIDE~\cite{microsoftcorporationSTRIDEThreatModel2009} is a widely adopted threat modelling framework.
It is particularly suitable for stakeholders with limited security expertise and has been applied to container and microservice architectures~\cite{wongThreatModelingSecurity2021, wongSecurityContainersThreat2023, tenevEnhancingSecurityMicroservice2019}.

These methods provide valuable foundations for ensuring risk completeness~\textit{(O2)} and supporting multifaceted risk representations~\textit{(O3)}.
However, they show limitations in terms of reproducibility~\textit{(O1)} and pervasive technical integration~\textit{(O4)}.

\paragraph{Formalised Rulesets for Vulnerability Detection}
Static code analysis~\cite{louridasStaticCodeAnalysis2006} and vulnerability scanning~\cite{werlingerPreparationDetectionAnalysis2010} rely on predefined rules to detect weaknesses in software artefacts.
These techniques are effective for identifying code-level issues, but often lack contextual awareness and integration with broader system configurations.
Vulnerability scanners for containers use rule graphs to assess security~\cite{doExplainingStaticAnalysis2022, doanDAVSDockerfileAnalysis2022, javedUnderstandingQualityContainer2021}.
Their design principles inform the structure of \textit{CSRO}, particularly in linking technical artefacts to risk indicators.

The Common Weakness Enumeration (CWE) and Common Vulnerabilities and Exposures (CVE) databases provide structured taxonomies of known security issues.
We aim to integrate these resources into \textit{CSRO} in the future, since are widely used in container security research~\cite{martinDockerEcosystemVulnerability2018, linMeasurementStudyLinux2018, wongSecurityContainersThreat2023}. 

Formalised rulesets offer a strong foundation for reproducible risk identification~\textit{(O1)} and technical integration~\textit{(O4)}.
However, evaluating completeness~\textit{(O2)} and adapting risks to specific contexts~\textit{(O3)} remains essential.

\paragraph{Cybersecurity Knowledge Graphs}
\textit{CSRO} is inspired by existing ontologies for security and risk modelling~\cite{adachCombinedSecurityOntology2022, salesCommonOntologyValue2018, kaloroumakisKnowledgeGraphCybersecurity2021}.
However, its instantiation as a knowledge graph does currently not integrate these resources directly since it primarily represents the risk assessment model observed in our case study.

Cybersecurity knowledge bases offer flexibility compared to ontologies, as they do not require a strict schema and can be constructed from heterogeneous data sources.
Examples include the MITRE ATT\&CK framework and domain-specific graphs for threat modelling~\cite{roweCoordinationModelAttack2023, sikosCybersecurityKnowledgeGraphs2023, wangSocialEngineeringCybersecurity2021}.

These approaches support cross-domain knowledge sharing and enable multifaceted risk representations~\textit{(O3)}.
They also contribute to reproducibility~\textit{(O1)}, but often lack mechanisms for context-dependent completeness~\textit{(O2)} and deep technical integration~\textit{(O4)}.

\section{Conclusion}
We present an ontology-based approach to security risk identification for container deployments in operational technology (OT) contexts.
Our approach integrates a modular knowledge model---formalised as the \textit{Container Security Risk Ontology (CSRO)}---with semantic reasoning and tool support to complement existing risk assessment practices.

\textit{CSRO} enables reproducible, interpretable, and technically integrated identification of security risks based on container artefacts and contextual assumptions.
We instantiate the ontology as a semantic knowledge graph and demonstrated its applicability through a case study of an industrial software platform.
The evaluation showed that our approach facilitates automated risk level calculation and the derivation of effective risk treatments.

Future work will focus on extending the scope of risk identification beyond container-level traits to include host system configurations and organisational measures.
We also aim to reduce the maintenance overhead of the knowledge base by clustering attack patterns and generalising risk calculation rules.
Furthermore, we plan to explore the integration of generative AI to enhance usability and support adaptive risk treatment recommendations.


\bibliographystyle{ACM-Reference-Format}
\bibliography{ASCA4IE_literature.bib}


\end{document}